\documentclass{appolb}
\usepackage{epsfig}
\usepackage{amsmath}
\usepackage{subfigure}
\usepackage{hyperref}
\begin{document}
\title{Non-uniform phases in a three-flavour 't Hooft extended Nambu-Jona--Lasinio model%
\thanks{Presented by JM at EEF70, the Workshop on Unquenched Hadron 
Spectroscopy: Non-Perturbative Models and Methods of QCD vs. Experiment, 
at the occasion of Eef van Beveren's 70th birthday.
}%
}
\author{
J. Moreira, B. Hiller,
\address{Centro de F\'{i}sica Computacional, Department of Physics,\\
University of Coimbra, P-3004-516 Coimbra, Portugal\\~}
\\
W. Broniowski,
\address{The H. Niewodnicza\'nski Institute of Nuclear Physics,\\
Polish Academy of Sciences, PL-31342 Krak\'ow, Poland\\
Institute of Physics, Jan Kochanowski University, PL-25406~Kielce, Poland\\~}\\
A. A. Osipov, A. H. Blin
\address{Centro de F\'{i}sica Computacional, Department of Physics,\\
University of Coimbra, P-3004-516 Coimbra, Portugal}
}
\maketitle
\begin{abstract}
The possible existence of non-uniform phases in cold dense quark matter in the
light quark sector ($u$, $d$, $s$) is addressed within the Nambu-Jona--Lasinio 
Model extended to include the flavour-mixing 't Hooft determinant. The effect 
of changes in the coupling strengths of the model is discussed. It is seen that 
the inclusion of the strange sector catalyses the appearance of the
non-uniform phases, extending the domain for their existence.
\end{abstract}

  
\section{Introduction}

It has been proposed long ago (for recent reviews 
see~\cite{Broniowski:2011}) that the non-trivial dynamics due 
to the attractive interaction of pions with quarks (or nucleons) allows for 
the existence of non-uniform phases in the low temperature and high chemical 
potential regime of the QCD phase diagram. As the 
sign problem affects lattice QCD at non-zero baryon density, the use of 
alternative approaches such as low energy models of the Nambu--Jona-Lasinio 
(NJL)~\cite{Nambu:1961} type, 
became a useful exploratory tool in the studies of strongly interacting matter.

In this talk we outline some results reported in detail in 
\cite{Moreira:2013ura}, pertaining to the study of the NJL model with the 
't Hooft determinant \cite{Hooft:1976} applied to the case where the light quark chiral 
condensates assume the shape of the chiral wave. We work
at zero temperature and in the chiral limit for the $u$ and $d$ sector, while 
the $s$ quark has a non-vanishing current mass.

\section{The model}

The chiral wave \textit{ansatz} proposed in \cite{Dautry:1979bk}, 
with the corresponding quark orbitals and energy levels, forms a self 
consistent solution of the Euler-Lagrange equations of the model in the chiral 
limit. They have the form
\begin{align}
\label{ansatz}
\langle\overline{\psi_l}\psi_l\rangle&=\frac{h_l}{2} \mathrm{cos}(\boldsymbol{q}\cdot\boldsymbol{r}), \quad
\langle\overline{\psi_l} i \gamma_5\tau_3\psi_l\rangle=\frac{h_l}{2} \mathrm{sin}(\boldsymbol{q}\cdot\boldsymbol{r}),\nonumber\\ 
E^{\pm}&=\sqrt{M^2+p^2+\frac{q^2}{4}\pm\sqrt{\left(\boldsymbol{p}\cdot\boldsymbol{q}\right)^2+M^2 q^2}},
\end{align}
where $\tau_3$ is the Pauli matrix acting in the isospin space, $M$ is the quark
dynamical mass, $\boldsymbol{p}$ denotes the momentum of the quark, and 
$\boldsymbol{q}$ is the wave vector of the condensate modulation. We choose the 
$z$-axis to coincide with $\boldsymbol{q}$ (note that the $E^-$ branch has a 
lower energy than $E^+$). For the $s$ quark a uniform condensate background 
is considered.

The application of techniques of Ref.~\cite{Osipov:2003xu} yields the 
thermodynamic potential of the model in the mean field approximation as
\begin{align}
\label{Omega}
\Omega=&V_{st}+\frac{N_c}{8\pi^2}\left(J_{-1}(M_u,\mu_u,q)+J_{-1}(M_d,\mu_d,q)+J_{-1}(M_s,\mu_s,0)\right)\nonumber\\
V_{st}=&\frac{1}{16}\left.\left(4G\left(h_u^2+h_d^2+h_s^2\right)+\kappa h_u h_d h_s\right)\right|^{M_i}_0,
\end{align}
where $h_i$ ($i=u,d,s$) are twice the quark condensates. The integrals $J_{-1}$ stem from the fermionic path integral over the quark
bilinears which appear after bosonization, while $V_{st}$ corresponds to the 
stationary phase contribution to the integration over the auxiliary bosonic 
fields. 
The NJL coupling strength is $G$, while $\kappa$ is the 
OZI-violating 't Hooft determinant coupling. From the value evaluated at the 
dynamical masses $M$, a subtraction of its value evaluated at $M=0$ is made 
\cite{Hiller:2008nu} (denoted by the $|^M_0$ symbol in Eq.~(\ref{Omega})). 
We use a regularized kernel corresponding to two Pauli-Villars 
subtractions in the integrand \cite{Osipov:1985}, 
namely 
$\rho\left(s\Lambda^2\right)=1-(1+s\Lambda^2)\mathrm{exp}(-s\Lambda^2)$. The 
Dirac and Fermi sea contributions, $J^{vac}_{-1}$ and $J^{med}_{-1}$, can be 
written explicitly as
\begin{align}
J_{-1}=&J^{vac}_{-1}+J^{med}_{-1},\nonumber\\
J^{vac}_{-1}=&\int\frac{\mathrm{d}^4 p_E}{(2\pi)^4}\int^\infty_0 \frac{\mathrm{d}s}{s}\rho\left(s\Lambda^2\right)8\pi^2e^{-s\left(p_ {0\,E}^2+p_\perp^2\right)}\nonumber\\
&\left.\left(e^{-s\left(\frac{q}{2}+\sqrt{M^2+p_z^2}\right)^2}+e^{-s\left(\frac{q}{2}-\sqrt{M^2+p_z^2}\right)^2}\right)\right|^{M,q}_{0,0},\nonumber\\
J^{med}_{-1} =&-\int\frac{\mathrm{d}^3p}{(2\pi)^3}8\pi^2T\left.\left(\mathcal{Z}^+_++\mathcal{Z}^+_-+\mathcal{Z}^-_++\mathcal{Z}^-_-\right)\right|^{M,q}_{0,0}+C(T,\mu),\nonumber\\
\mathcal{Z}^\pm_\pm =&\mathrm{log}\left(1+e^{-\frac{E^\pm\mp\mu}{T}}\right)-\mathrm{log}\left(1+e^{-\frac{E_\Lambda^\pm\mp\mu}{T}}\right)-
\frac{\Lambda^2}{2T E_\Lambda^\pm}\frac{e^{-\frac{E_\Lambda^\pm\mp\mu}{T}}}{1+e^{-\frac{E_\Lambda^\pm\mp\mu}{T}}},\nonumber\\
C(T,\mu)=&\int\frac{\mathrm{d}^3p}{(2\pi)^3}16\pi^2T
\, 
\mathrm{log}\left(\left(1+e^{-\frac{|\boldsymbol{p}|-\mu}{T}}\right)\left(1+e^ { 
-\frac{|\boldsymbol{p}|+\mu}{T}}\right)\right)
\end{align}
where $E_\Lambda^{\pm}=\sqrt{\left(E^{\pm}\right)^2+\Lambda^2}$. The 
$|^{M,q}_{0,0}$ notation refers to the subtraction of 
the same quantity evaluated for $M=0$ and $q=0$, which is done to set the 
zero of the potential at a uniform gas of massless quarks.
The superscript $\pm$ in the definition of $\mathcal{Z}$ refers to the energy 
branch, whereas the subscript refers to the sign in front of the chemical 
potential in the exponent. The $C(T,\mu)$ term is needed for thermodynamic 
consistency \cite{Hiller:2008nu}.
The minimization of the $\Omega$ with respect to $M$ and $q$ has to be done self-consistently with the resolution of the following stationary phase equations (where $m_i$ stands for the current masses of the quarks):
\begin{align}
\label{StaEq}
\left\{
\begin{array}{l}
m_u-M_u=G h_u +\frac{\kappa}{16}h_d h_s\\
m_d-M_d=G h_d +\frac{\kappa}{16}h_u h_s\\
m_s-M_s=G h_s +\frac{\kappa}{16}h_u h_d
\end{array}	
\right..
\end{align}

\section{Results}

In our study we chose the parameters to reproduce a reasonable value 
for the vacuum dynamical mass of the light quarks ($M_l=300~\mathrm{MeV}$). The 
current masses were set to $m_u=m_d=0$, $m_s=186~\mathrm{MeV}$, the remaining 
three parameters of the model ($G$, $\kappa$ and $\Lambda$) can then reduced to 
two: the 't Hooft coupling strength, $\kappa$, and the dimensionless curvature, 
$\tau=N_c G \Lambda^2/(2\pi^2)$. In the chiral limit $\tau=1$ is the critical 
value above which dynamical chiral symmetry breaking occurs as a crossover for 
$1<\tau<1.23$ and as a first order transition for higher values.

With no OZI-violating term, the light and strange sectors are 
decoupled: two independent first order transitions occur for the  light and 
strange sector, with the latter shifted to higher $\mu$ due to the current 
$s$-quark mass (both occur at values of $\mu$ close to the vacuum 
dynamical mass $M^{\rm vac}_i$).
At high enough $\mu$, the energetically 
favourable solution is always a modulated one, with a non-vanishing $q$ -- the 
non-uniform phase develops. Asymptotically, this solution goes to 
$\lim_{\mu\rightarrow\infty}\left\{h,q\right\}=\{0,2\mu\}$ and becomes 
degenerate with the trivial one. Above $1.23<\tau<1.53$, a window for
energetically favourable finite-$q$ solutions appears in the vicinity of the 
first-order transition encompassing it; the transition is not to a vanishing 
condensate, but $\left\{h_l,0\right\}\rightarrow\left\{h_l^\prime,q\right\}$. 
The window ends with the light condensate going continuously to zero. 
For $\tau>1.53$ this windows extends and merges with the 
one at higher chemical potentials. 
\begin{figure}[htb]
\vspace{-5mm}
\centerline{\includegraphics[width=\textwidth]{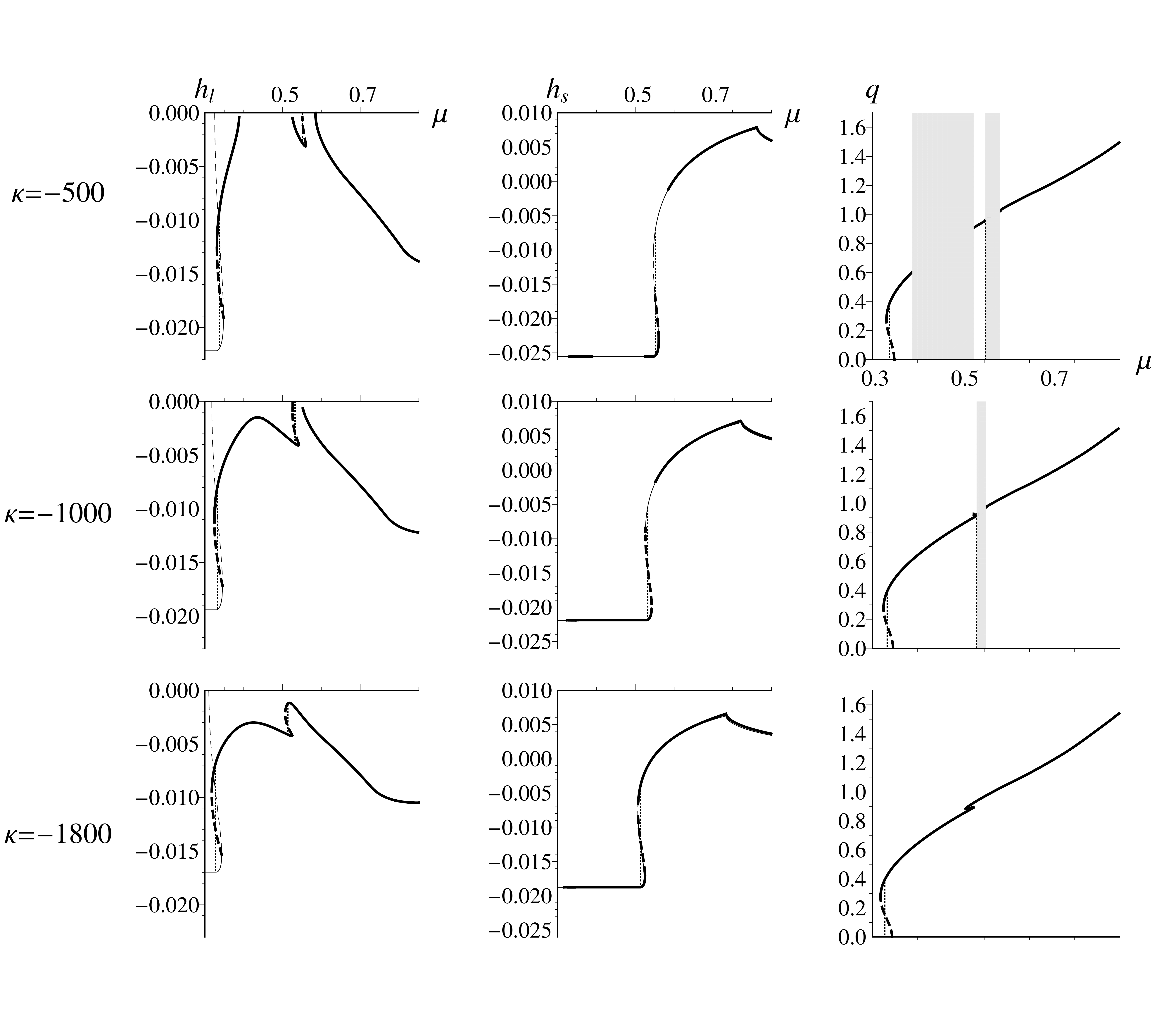}}
\vspace{-5mm}
\caption{The $\mu$-dependence of $h_l$, $h_s$ and $q$. Each row corresponds
to a different value of the 't Hooft coupling strength
($\left[\kappa\right]=\mathrm{GeV}^5$). Thicker lines denote the 
finite-$q$ solutions. In the grey regions the value of $q$ is undetermined, 
since $h_l=0$.}
\label{grafpainelNJLHSU3}
\end{figure}

\begin{figure}[htb]
\vspace{-5mm}
\centerline{%
\subfigure[]{\label{grafSolhlNJLHtau14K500zoomI}
\includegraphics[width=0.25\textwidth]{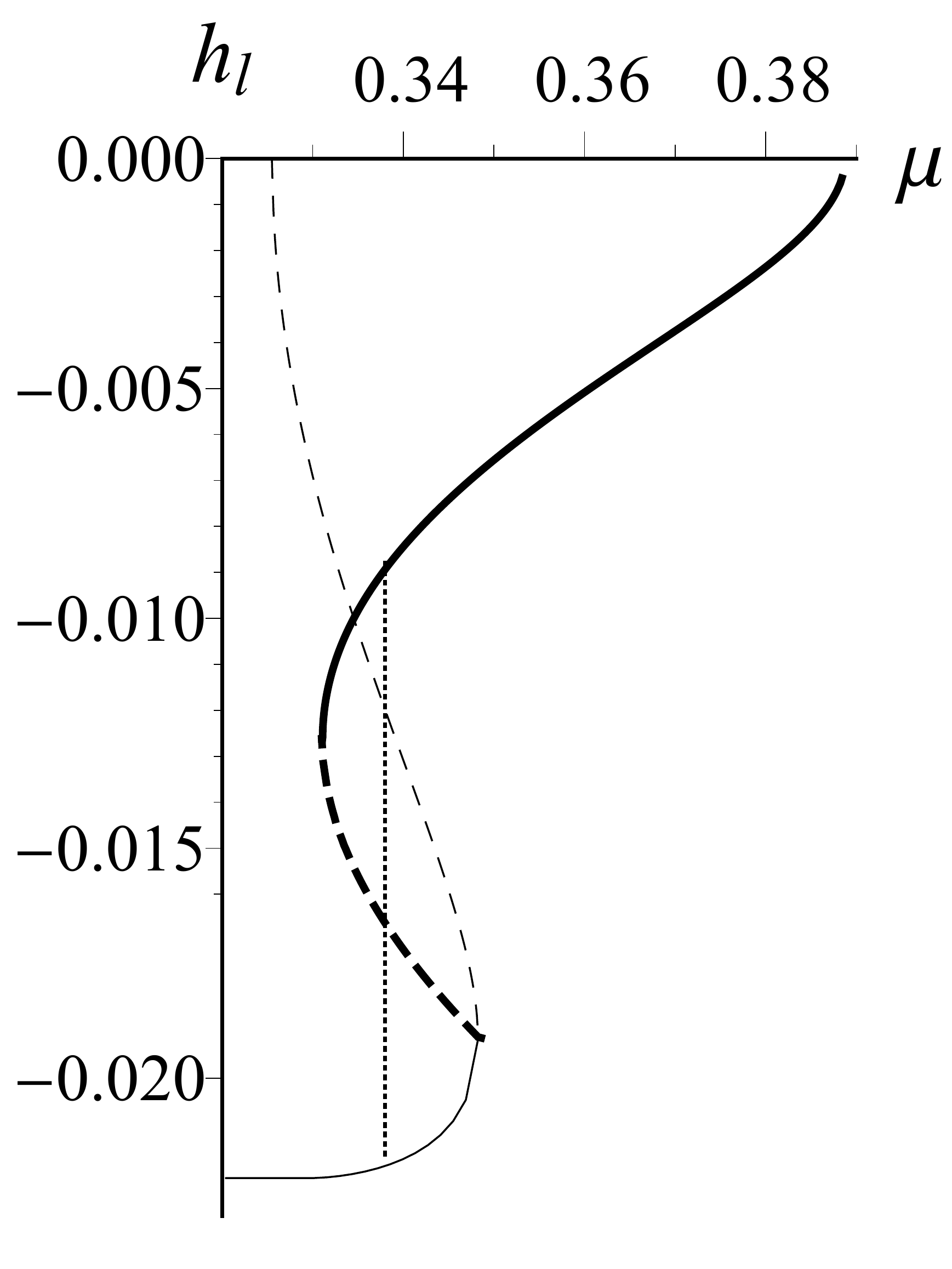}}
\subfigure[]{\label{grafSolhsNJLHtau14K500zoomI}
\includegraphics[width=0.25\textwidth]{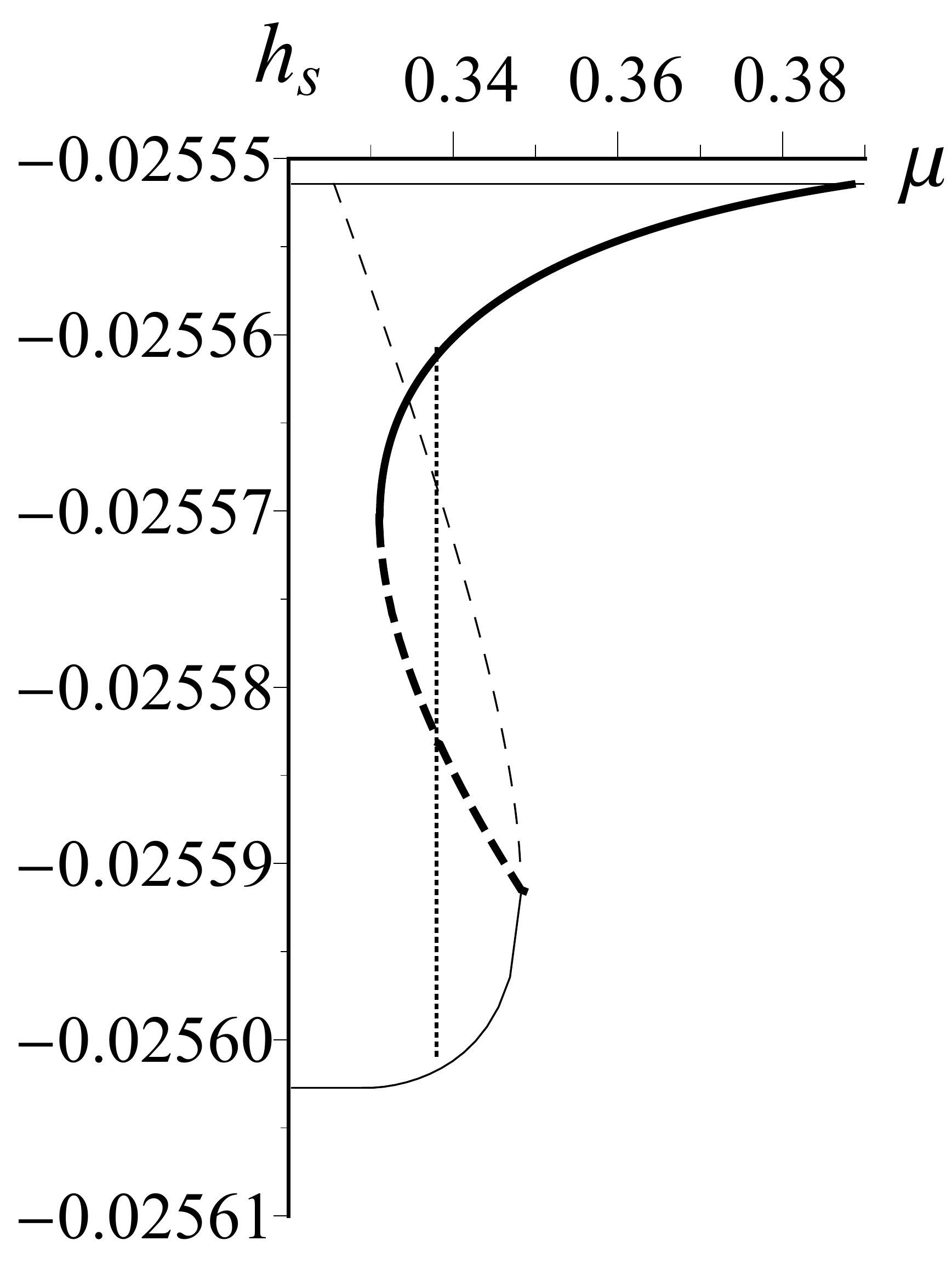}}
\subfigure[]{\label{grafSolhlNJLHtau14K500zoomII}
\includegraphics[width=0.25\textwidth]{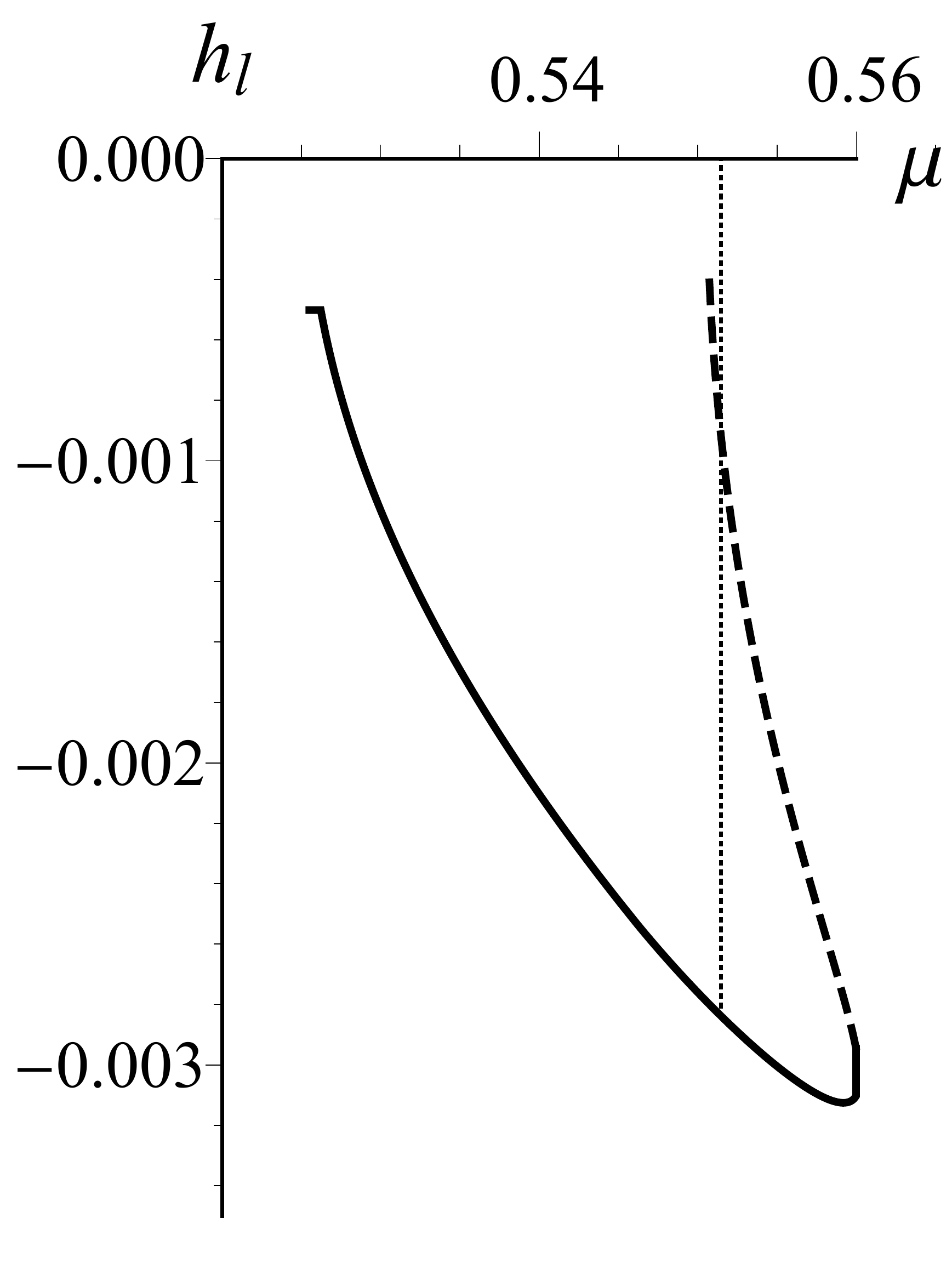}}
\subfigure[]{\label{grafSolhsNJLHtau14K500zoomII}
\includegraphics[width=0.25\textwidth]{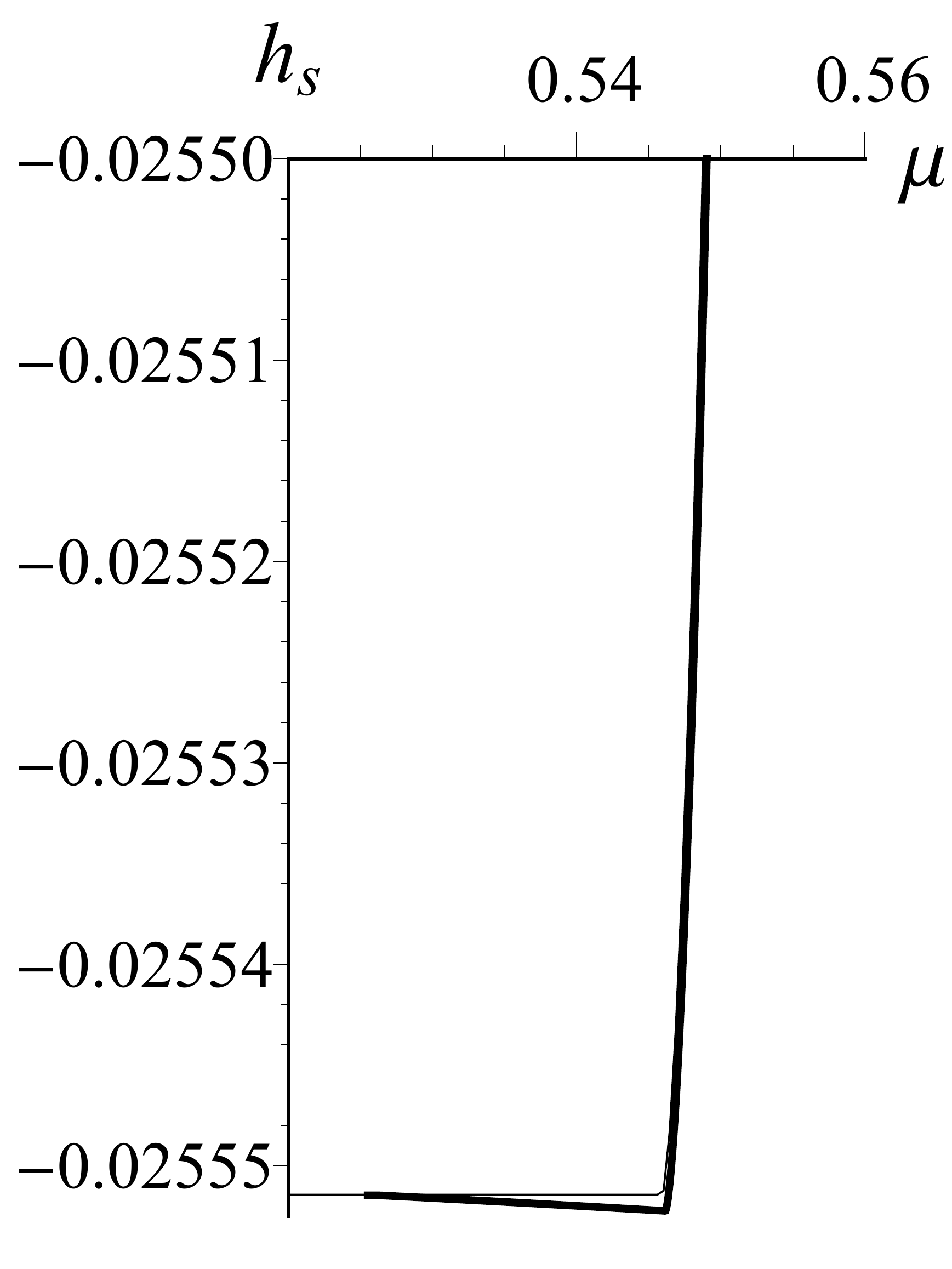}}}
\caption{Zoom for the chemical potential dependence of the chiral condensate 
solutions (thicker lines refer to finite-$q$ solutions) near the transitions 
(marked by the vertical dotted lines), case of 
$\kappa=-500~\mathrm{GeV}^{-5}$.}
\label{SolhlsNJLHtau14K500zoom}
\end{figure}

\begin{figure}
\vspace{-5mm}
\center
\subfigure[]{\label{grafSolqNJLHtau14K500zoom}\includegraphics[
width=0.25\textwidth]{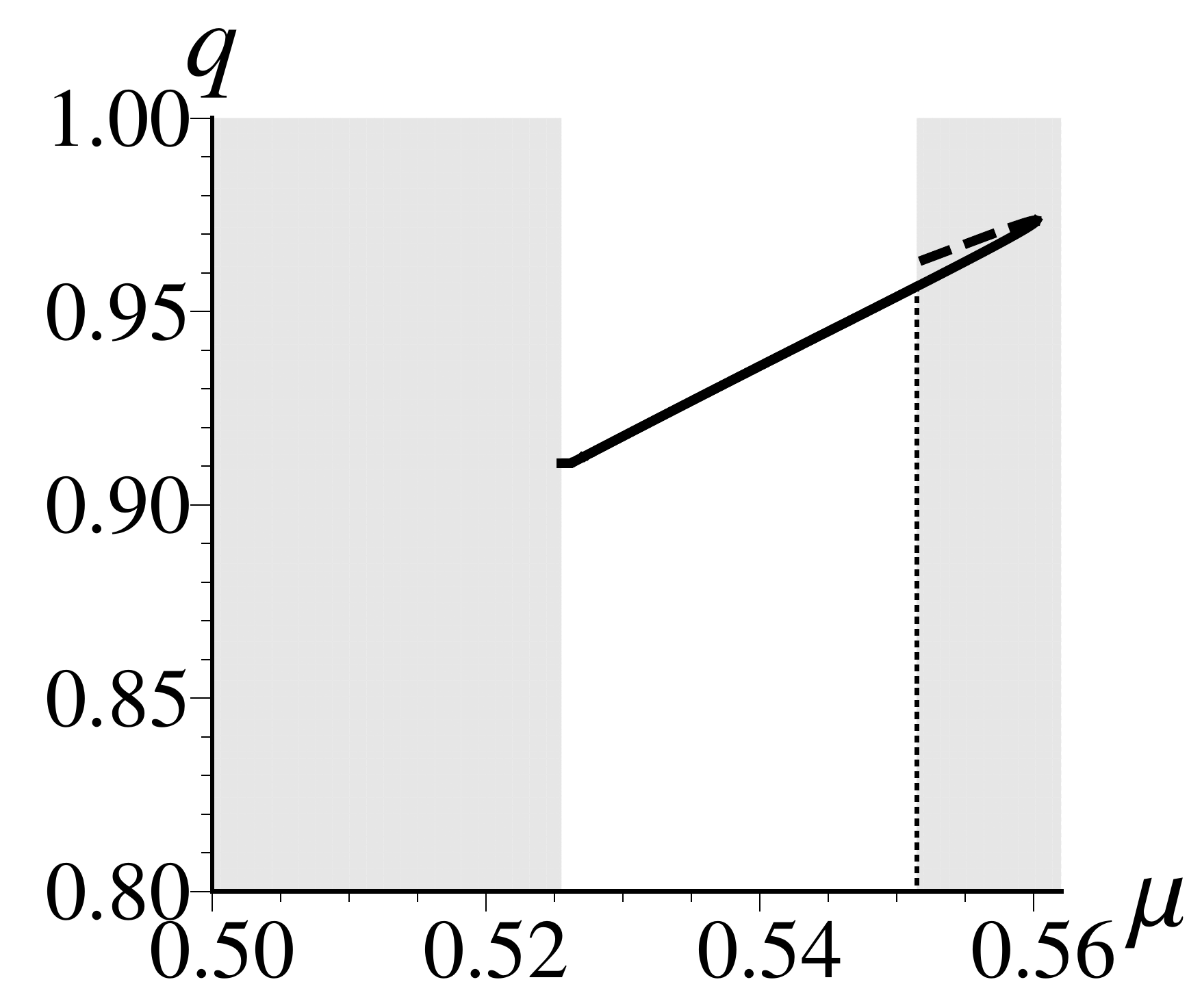}}
\subfigure[]{\label{grafSolqNJLHtau14K1000zoom}\includegraphics[
width=0.25\textwidth]{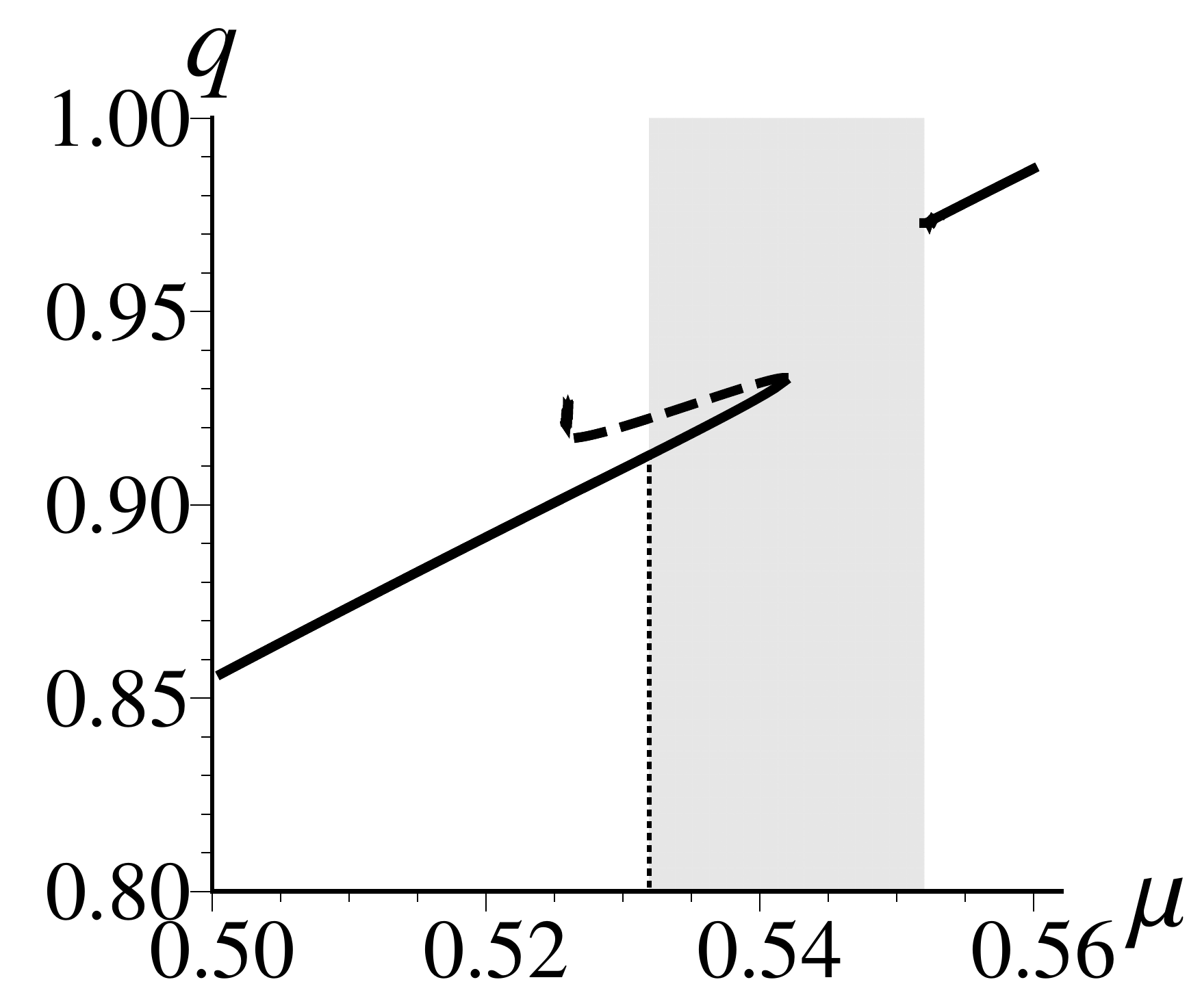}}
\subfigure[]{\label{grafSolqNJLHtau14K1800zoom}\includegraphics[
width=0.25\textwidth]{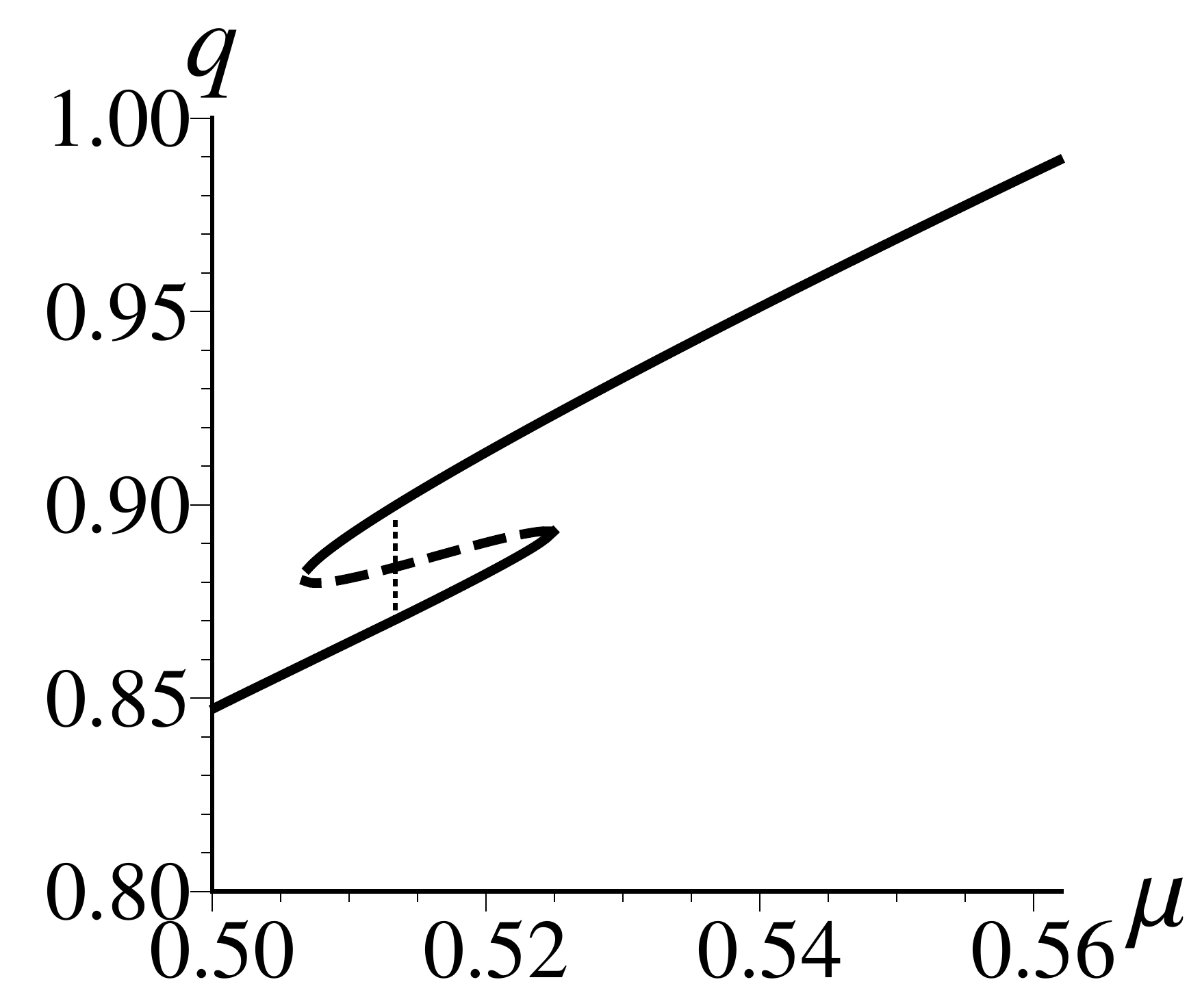}}
\caption{Zoom of the solutions for $q$ in the chemical potential window close to 
$M^{\rm vac}_s$ for $\kappa=-500,-1000$, and $-1800~\mathrm{GeV}^5$ (from left 
to right).
}
\label{grafSolNJLHtau14K1000}
\end{figure}

\begin{figure}
\vspace{-5mm}
\center
\subfigure[]{\label{grafPotQuimCritVsKNJLHCWtau14shade}\includegraphics[
width=0.35\textwidth]{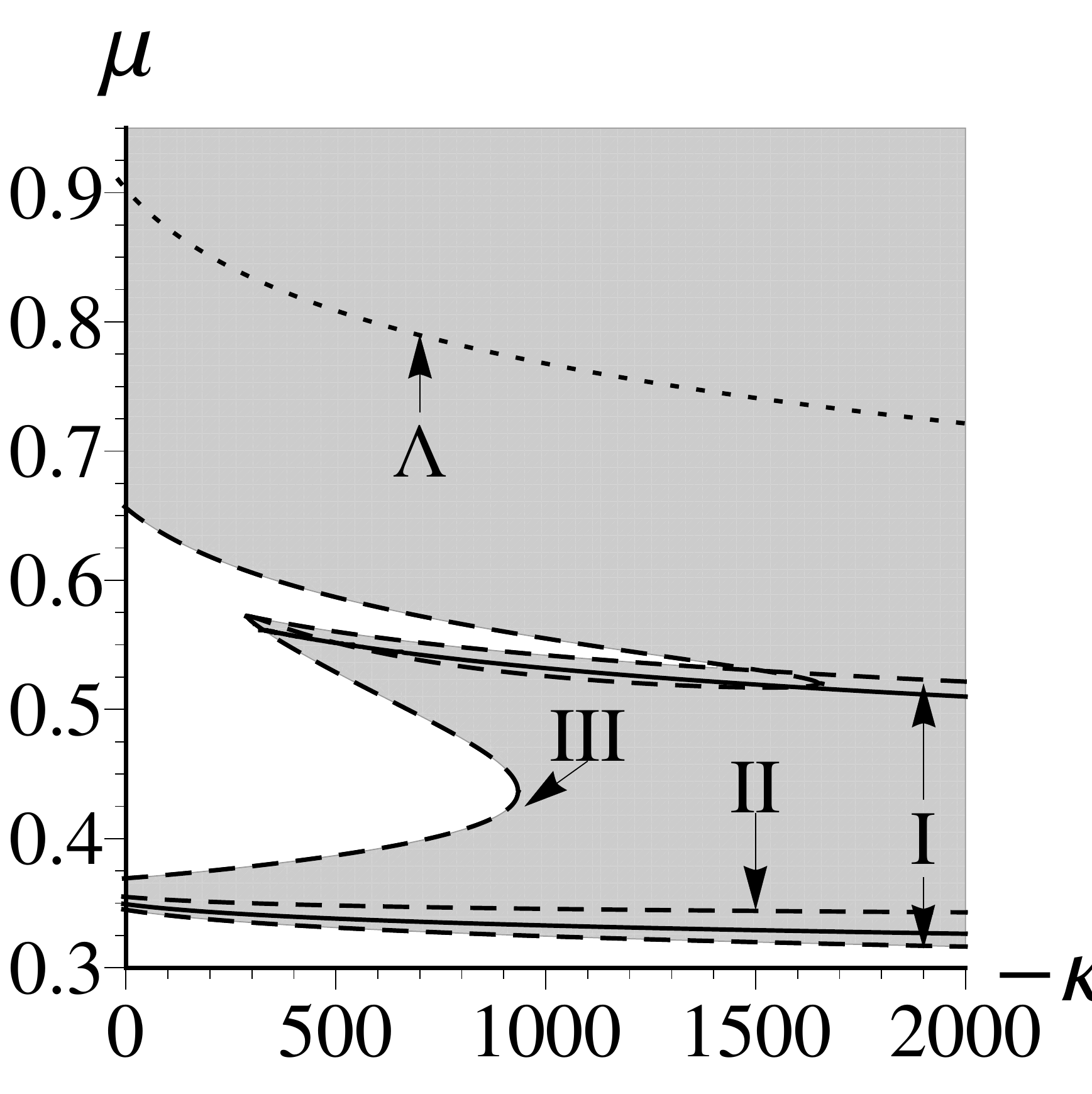}}
\subfigure[]{\label{grafAuxPotQuimSolNJLHCW}\includegraphics[
width=0.35\textwidth ]{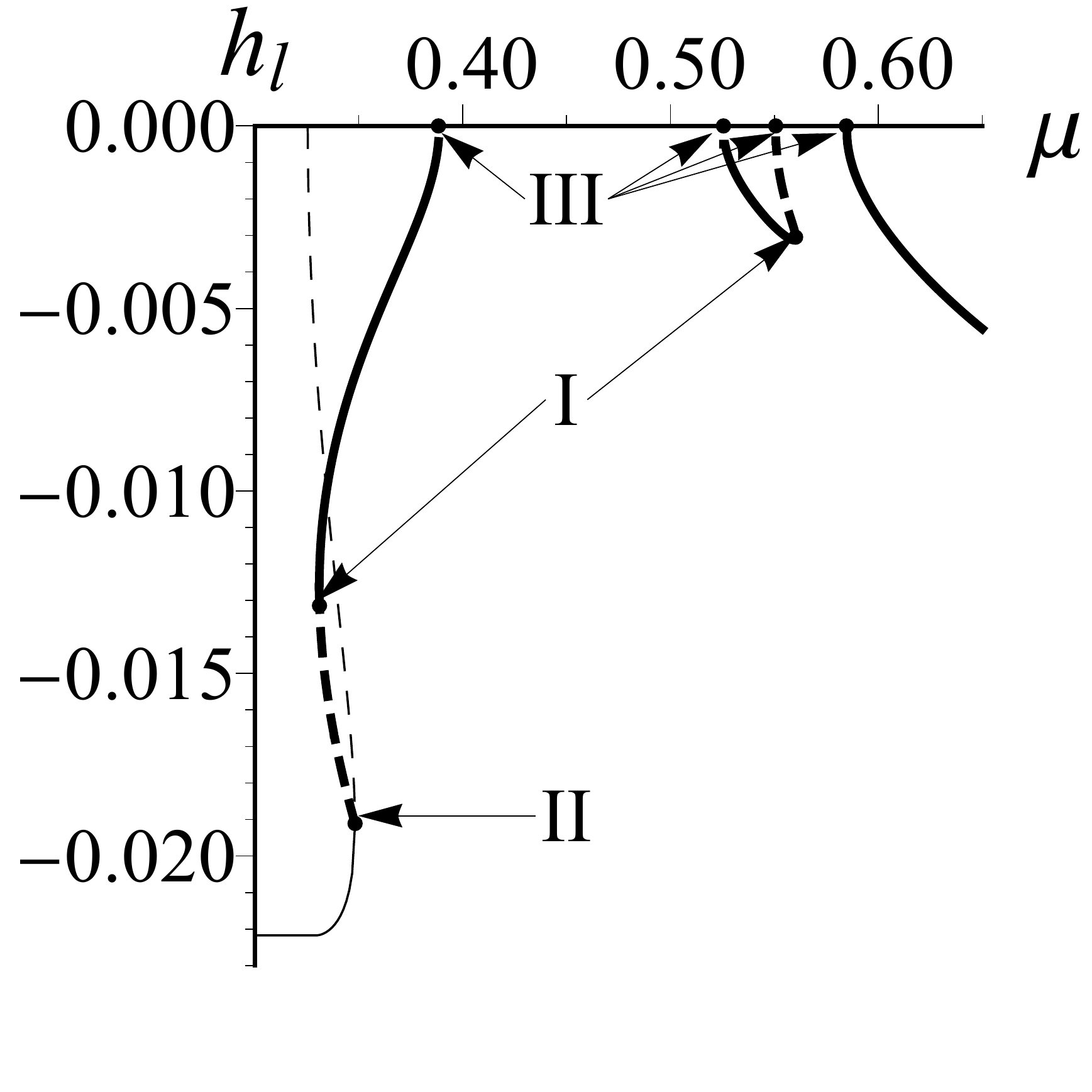}}
\caption{Critical chemical potentials (in $\mathrm{GeV}$) as functions of 
$\kappa$ ($[\kappa]=\mathrm{GeV}^5$). The upper dotted line corresponds to the 
cut-off $\Lambda$. The chemical potential of the first order transitions are 
marked with the full black lines. Dashed lines indicate the borders of the 
region where the finite-$q$ solutions exist. We distinguish between three types 
of critical chemical potentials and an example for this distinction in the 
$\kappa=-500~\mathrm{GeV}^{-5}$ case appears in
Fig.~\ref{grafAuxPotQuimSolNJLHCW} (type I with finite $q$ and $h_l$, type II with finite $h_l$ and vanishing $q$, and type III with finite $q$ but vanishing $h_l$).}
\label{muCritVsK}
\end{figure}

Turning on flavour mixing couples the $u-d$ and strange 
sectors. At fixed curvature, we find several different scenarios depending 
on the value of $\kappa$, as seen in Fig.~\ref{grafpainelNJLHSU3} where we
show the $\tau=1.4$ case. For $-\kappa>290~\mathrm{GeV}^{-5}$ a new 
solution branch appears, with a shark-fin  shape for $h_l$  in the vicinity of 
$\mu\sim M^{\rm vac}_s$ (see first row of Fig.~\ref{grafpainelNJLHSU3}). 
There are therefore three intervals of $\mu$ where an energetically favourable 
finite-$q$ solution exists. They are delimited by two first order transitions 
and three-crossovers (the crossovers correspond to the disappearance/emergence 
of a non-vanishing $h_l$). The two first order transitions occur slightly 
below (for the one taking place near $M_l^{\rm vac}$) and slightly above (near 
$M_s^{\rm vac}$), thus excluding the occurrence of the $q=0$ transitions. A 
zoom of the behaviour of the chiral condensates near the transitions can be 
seen in Fig.~\ref{SolhlsNJLHtau14K500zoom}. For $-\kappa>935~\mathrm{GeV}^{-5}$ 
the first two $\mu$ windows with finite-$q$ solutions merge, 
resulting in the disappearance of the corresponding crossover transitions, as 
can be seen in the second row of Fig.~\ref{grafpainelNJLHSU3}.  With 
$-\kappa>1660~\mathrm{GeV}^{-5}$ the last transition to a vanishing $q$ solution 
does not occur and is substituted by a first-order transition between the two
phases with finite $q$ (third row of Fig.~\ref{grafpainelNJLHSU3} and 
Fig. \ref{grafSolNJLHtau14K1000}). The dependence on $\kappa$ of these critical 
chemical potentials can be seen in Fig.~\ref{muCritVsK}. 

We conclude that the flavour mixing acts as a 
catalyst for the emergence of globally stable inhomogeneous solutions in 
zero-temperature quark matter. 

This work has been supported by the Centro de F\'{i}sica Computacional of the University of Coimbra, Funda\c{c}\~ao para a Ci\^encia e Tecnologia, project: CERN/FP/116334/2010
and the grant SFRH/BPD/63070/2009/. This research is part of the EU Research
Infrastructure Integrating Activity Study of Strongly Interacting Matter 
(HadronPhysics3) under the 7th Framework Programme of EU, Grant Agreement No. 
283286. WB acknowledges the support of the Polish National Science 
Centre, grants DEC-2011/01/B/ST2/03915 and DEC-2012/06/A/ST2/00390.


\begin{thebibliography}{99}
\bibitem{Broniowski:2011} W. Broniowski, Acta Phys.Pol.Supp. {\bf 4} (2012) 631.  
\bibitem{Buballa:2014} M. Buballa, S. Carignano, arXiv:1406.1367 [hep-ph].
\bibitem{Nambu:1961}Y. Nambu, Phys. Rev. Lett. {\bf 4} (1960) 380; Phys. Rev. {\bf 122}, 
   (1961)  345; {\bf 124} (1961) 246; V. G. Vaks and A. I. Larkin, 
 Sov. Phys. JETP {\bf 13} (1961) 192.
\bibitem{Moreira:2013ura} J. Moreira et al., Phys. Rev. D {\bf 89} (2014) 3, 036009. 
\bibitem{Hooft:1976} G. 't Hooft, Phys. Rev. D {\bf 14} (1976) 3432.
\bibitem{Dautry:1979bk} F. Dautry, E. Nyman, Nucl. Phys. A {\bf 319} (1979) 323.
\bibitem{Osipov:2003xu} A. A. Osipov, B. Hiller, Eur. Phys. J. C {\bf 35} (2004) 223.
\bibitem{Hiller:2008nu} B. Hiller et al., Phys. Rev. D {\bf 81} (2010) 116005.
\bibitem{Osipov:1985} A. A. Osipov, M. K. Volkov, Sov. J. Nucl. Phys. {\bf 41}:3 (1985) 500.
\end{thebibliography}
\end{document}